\documentclass[preprint]{vgtc}
\usepackage[dvipsnames]{xcolor}
\usepackage{listings}
\usepackage{glslListings}
\usepackage{kotlinListings}
\usepackage{courier}
\ifpdf%                                % if we use pdflatex
  \pdfoutput=1\relax                   % create PDFs from pdfLaTeX
  \usepackage{graphicx}                % allow us to embed graphics files
  \usepackage{hyperref}
  \DeclareGraphicsExtensions{.pdf,.png,.jpg,.jpeg} % for pdflatex we expect .pdf, .png, or .jpg files
\else%                                 % else we use pure latex
  \usepackage{graphicx}                % allow us to embed graphics files
  \DeclareGraphicsExtensions{.eps}     % for pure latex we expect eps files
\fi%

\graphicspath{{figures/}{pictures/}{images/}{./}} % where to search for the images

\usepackage{microtype}                 % use micro-typography (slightly more compact, better to read)
\PassOptionsToPackage{warn}{textcomp}  % to address font issues with \textrightarrow
\usepackage{textcomp}                  % use better special symbols
\usepackage{mathptmx}                  % use matching math font
\usepackage{times}                     % we use Times as the main font
\usepackage{cite}                      % needed to automatically sort the references
\usepackage{tabu}                      % only used for the table example
\usepackage{booktabs}                  % only used for the table example
\usepackage{enumitem}
\usepackage{mathpazo}

\newenvironment{enumerate*}%
  {\begin{enumerate}[label=\subscript{G}{{\arabic*}}]%
    \setlength{\itemsep}{1.5pt}%
    \setlength{\parskip}{2pt}}%
  {\end{enumerate}}

\newcommand{\subscript}[2]{$#1 _ #2$}
\newcommand{\revised}[1]{{#1}}
\usepackage[anythingbreaks]{breakurl}
\onlineid{1060}
\ieeedoi{10.1109/VISUAL.2019.8933605}

\vgtccategory{Research}

\vgtcinsertpkg

\title{scenery: Flexible Virtual Reality Visualization on the Java VM}

\author{Ulrik~G\"{u}nther\thanks{e-mail: guenther@mpi-cbg.de}\\ 
    \parbox{1.8in}{\scriptsize \centering CASUS, G\"orlitz\\ Technische Universit\"at Dresden \\ Center for Systems Biology Dresden \\ MPI-CBG, Dresden  }
\and        Tobias Pietzsch\\%\thanks{e-mail: pietzsch@mpi-cbg.de}\\
    \parbox{1.8in}{\scriptsize \centering Center for Systems Biology Dresden \\ MPI-CBG, Dresden}
\and        Aryaman Gupta\\%\thanks{e-mail: argupta@mpi-cbg.de}\\
    \parbox{1.8in}{\scriptsize \centering Technische Universit\"at Dresden \\ Center for Systems Biology Dresden \\ MPI-CBG, Dresden }
\vspace{1.5em}
\and        Kyle~I.S.~Harrington\\%\thanks{e-mail: kharrington@uidaho.edu}\\
    \parbox{1.8in}{\scriptsize \centering University of Idaho \\
    Howard Hughes Medical Institute, Janelia Research Campus }
\and        Pavel~Tomancak\\%\thanks{e-mail: tomancak@mpi-cbg.de}\\
    \parbox{1.8in}{\scriptsize \centering MPI-CBG, Dresden \\
    IT4Innovations, V\v{S}B - Technical University of Ostrava}
\and        Stefan Gumhold\\%\thanks{e-mail: stefan.gumhold@tu-dresden.de}\\
    \parbox{1.8in}{\scriptsize \centering Technische Universit\"at Dresden }
    \vspace{0.7cm}
\and        Ivo~F.~Sbalzarini\thanks{e-mail: ivos@mpi-cbg.de}\\
    \parbox{1.8in}{\scriptsize \centering Technische Universit\"at Dresden \\ Center for Systems Biology Dresden \\ MPI-CBG, Dresden }
    \vspace*{-0.5cm}
}

\abstract{Life science today involves computational analysis of a large amount and variety of data, such as volumetric data acquired by state-of-the-art microscopes, or mesh data from analysis of such data or simulations. Visualization is often the first step in making sense of data, and a crucial part of building and debugging analysis pipelines. It is therefore important that visualizations can be quickly prototyped, as well as developed or embedded into full applications. In order to better judge spatiotemporal relationships, immersive hardware, such as Virtual or Augmented Reality (VR/AR) headsets and associated controllers are becoming invaluable tools. In this work we introduce \emph{scenery}, a flexible VR/AR visualization framework for the Java VM that can handle mesh and \revised{large volumetric data}, containing multiple views, timepoints, and color channels. scenery is free and open-source software, works on all major platforms, and uses the Vulkan or OpenGL rendering APIs. We introduce scenery's main features and example applications, such as its \revised{use in VR for microscopy}, in the biomedical image analysis software Fiji, or for visualising agent-based simulations.
}% end of abstract

\CCScatlist{
  \CCScatTwelve{Human-centered computing}{Visu\-al\-iza\-tion}{Visu\-al\-iza\-tion systems and tools}{}{}
  \CCScatTwelve{Human-centered computing}{Virtual reality}{}{}
}
\firstsection{Introduction}

\begin{document}
\maketitle

Recent innovations in biology, like lightsheet microscopy \cite{huisken2004optical}, or Serial Block-Face Scanning Electron Microscopy \cite{denk2004} are now making large, spatiotemporally complex volumetric data available. However, the data acquired by the microscope is only a means to an end: researchers need to extract results from it, and for that efficient tools are needed. This includes tools that not only enable the researcher to visualize their data, but to interact with it, and to enable them to use the tool in ways the original designer had not anticipated. 

For this purpose, we introduce \emph{scenery}, a flexible, open-source visualization framework for the Java Virtual Machine (JVM) that can handle mesh data (e.g. from triangulated surfaces), and multi-channel, multi-timepoint, multi-view volumetric data of \revised{large} size\footnote{Out-of-core data is stored in tiles, with $64$ bit tile indices, and each tile comprising up to $2^{31}$ voxels. Therefore the \revised{theoretical} limit for a single volume is $2^{94}$ voxels, \revised{roughly corresponding to a cube with $2.1$ billion voxels edge length, equal to $20000$ Yottabyte. The largest tested dataset was an 8TB multi-angle time-series, with 7GB per timepoint.}}. \revised{Our main contribution with scenery is to combine all of the following design goals into one reusable, open-source framework:}

\vspace*{-0.5\baselineskip}
\begin{enumerate*}
    \item \textbf{Virtual/Augmented Reality support}: The framework should make the use of VR/AR in an application possible with minimal effort. Distributed systems, such as CAVEs or Powerwalls, should also be supported.
    
    \item \textbf{Out-of-core volume rendering}: The framework should be able to handle datasets that do not fit into graphics memory and/or main memory, contain multiple channels, views, and timepoints. It should be possible to visualize multiple such datasets in a single scene.

    \item \textbf{User/Developer-friendly API}: The framework should have a simple API that makes only limited use of advanced features, such as generics, so the user/developer can quickly comprehend and customize it.
    
    \item \textbf{Cross-platform}: The framework should run on the major operating systems: Windows, Linux, and macOS.
    
    %\item \textbf{Free and open-source software}: The framework needs to be available to all researchers for free, and all %of its source code needs to be available for inspection and modification. 
    %Open-source software can boost reproducibility (reference? %https://www.nature.com/articles/nmeth.2082)
    %) and makes customisations easier.
    
    \item \textbf{JVM-native and embeddable}: The framework should run natively on the JVM, and be embeddable, such that it can be used in popular biomedical image analysis tools like Fiji \cite{schindelin2012fiji,Rueden:2017ij2}, Icy \cite{Chaumont:2012icy}, and KNIME \cite{Berthold:2008kni}.
\end{enumerate*}

\section{Related work}

A particularly popular \emph{framework} in scientific visualization is \emph{VTK} \cite{Hanwell:2015iv}: VTK offers rendering of both geometric and volumetric data, using an OpenGL 2.1 renderer. However, VTK's complexity has also grown over the years and its API is becoming more complex, making it difficult to change internals without breaking existing applications ($G_3$). A more recent development is \emph{VTK.js}
, which brings VTK to web browsers. \emph{ClearVolume} \cite{Royer:2015tg} is a visualization toolkit tailored to high-speed, volumetric microscopy and supports multi-channel/multi-timepoint data, but focuses solely on volumetric data and does not support VR/AR. \emph{MegaMol} \cite{grottel2014} is a special-purpose framework focused on efficient rendering of a large number of discrete particles that provides a thin abstraction layer over the graphics API for the developer. \emph{3D Viewer} \cite{Schmid:2010gm} does general-purpose image visualization tasks, and supports multi-timepoint data, but no out-of-core volume rendering, or VR/AR. 

In \emph{out-of-core rendering} (OOCR), the rendering of volumetric data that does not fit into main or graphics memory, existing software packages include \emph{Vaa3D/Terafly} \cite{peng2014extensible,Bria:2016fl}, which is written with applications like neuron tracing in mind, and \emph{BigDataViewer} \cite{pietzsch2015bigdataviewer}, which performs by-slice rendering of large datasets, powered by the ImgLib2 library\cite{Pietzsch:2012img}. The \emph{VR neuron tracing tool} \cite{Usher:2017bda} supports OOCR, but lacks support for multiple timepoints and is not customizable. \emph{Inviwo} \cite{jonsson2019} supports OOCR and interactive development, but does not support overlaying multiple volumetric datasets in a single view.

In the field of biomedical image analysis, various \emph{commercial packages} exist: \emph{Arivis}, \emph{Amira}, and \emph{Imaris}\footnote{\raggedright See  \href{https://www.arivis.com/en/imaging-science/imaging-science}{arivis.com/en/imaging-science/imaging-science}, \href{https://www.fei.com/software/amira/}{fei.com/software/amira/}, and \href{https://imaris.oxinst.com/}{imaris.oxinst.com}}, and \emph{syGlass} \cite{pidhorskyi2018} support out-of-core rendering, and are scriptable by the user. Arivis, Imaris, and syGlass offer rendering to VR headsets, while Amira can run on CAVE systems. Imaris provides limited Fiji and Matlab integration. Due to being closed-source, the flexibility of these packages is ultimately limited (\textit{e.g.}, changing rendering methods, or adding new input devices).

\section{scenery}

\begin{figure}[tb]
 \centering
 \includegraphics[width=\columnwidth]{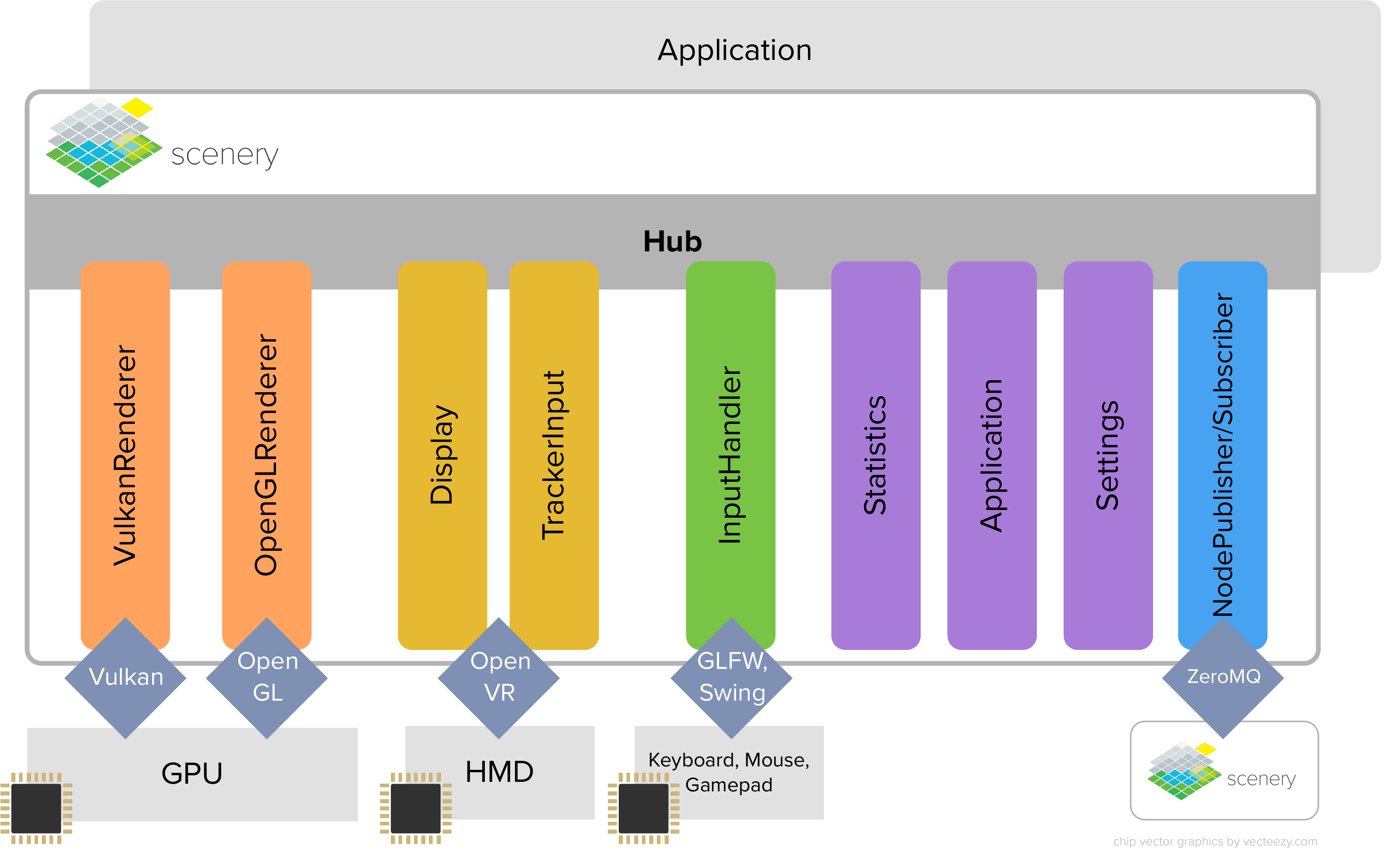}
 \caption{Overview of scenery's architecture.\label{fig:SceneryArchitecture}}
\end{figure}

With \emph{scenery}, we provide a flexible framework for developing visualization prototypes and applications, on systems ranging from desktop screens, VR/AR headsets (like the Oculus Rift or HTC Vive), to distributed setups. scenery is written in Kotlin, a language for the JVM that requires less boilerplate code and has more functional constructs than Java itself. This increases developer productivity, while maintaining 100\% compatibility with existing Java code. scenery runs on Windows, Linux, and macOS ($G_4$). scenery uses the low-level Vulkan API for fast and efficient rendering, and can fall back to an OpenGL 4.1-based renderer\footnote{The Vulkan renderer uses the LWJGL Vulkan bindings (see \href{https://lwjgl.org}{lwjgl.org}), while the OpenGL renderer uses JOGL (see \href{http://www.jogamp.org}{jogamp.org}).}.

scenery is designed around two concepts: A scene graph for the scene organisation into nodes, and a hub organizing all subsystems --- e.g. rendering, input, statistics --- and enables communication between them. scenery's application architecture is depicted in Fig.~\ref{fig:SceneryArchitecture}. scenery's subsystems are only loosely coupled, meaning they can work fully independent of each other. The loose coupling enables isolated testing of the subsystems, and thereby we can reach 65\% code coverage at the moment (the remaining 35\% is mostly code that requires additional hardware and is therefore harder to test in an automated manner).

\section{Highlighted features}

\subsection{Realtime rendering on the JVM --- $G_5$}

Historically, the JVM has not been the go-to target for realtime rendering: For a long time, the JVM had the reputation of being slow and memory-hungry. However, since the HotSpot VM has been introduced in Java 6, this is less true, and state-of-the-art just-in-time compilers like the ones used in Java 12 have become very good at generating automatically vectorized code\footnote{For this project, we have measured the timings of performance-critical parts of code, such as 4x4 matrix multiplication. Compared to hand-tuned, vectorized AVX512 code, the native code generated by the JVM's JIT compiler is about a factor of 3-4 slower.}. The JVM is widely used, provides excellent dependency management via the \emph{Maven} or \emph{Gradle} build tools, and efficient, easy-to-use abstractions for, \textit{e.g.}, multithreading or UIs on different operating systems. Additionally, with the move to low-overhead APIs like Vulkan, pure-CPU performance is becoming less important. In the near future, \emph{Project Panama}\footnote{See \href{https://openjdk.java.net/projects/panama/}{openjdk.java.net/projects/panama}.} will introduce JVM-native vectorization primitives to support CPU-heavy workloads. These primitives  will work in a way similar to those provided by .NET.

Another convenience provided by the JVM is scripting: Via the JVM's scripting extensions, scenery can be scripted using its REPL with third-party languages like Python, Ruby, and Clojure. In the future,  \emph{GraalVM}\footnote{See \href{https://www.graalvm.org}{graalvm.org}.} will enable polyglot code on the JVM, e.g. by ingesting LLVM bytecode directly\cite{bonetta2018}. scenery has already been tested with preview builds of both GraalVM and Project Panama.

\subsection{Out-of-core volume rendering --- $G_2$}
scenery supports volume rendering of multiple, potentially overlapping volumes that are placed into the scene via arbitrary affine transforms.
For out-of-core direct volume rendering of large volumes ($G_2$) we develop and integrate the \emph{BigVolumeViewer} library, which builds on the pyramidal image data structures and in-memory caching of large image data from BigDataViewer \cite{pietzsch2015bigdataviewer}.
We augment this by a GPU cache tier for volume blocks, implemented using a single large 3D texture.
This cache texture is organized into small (\textit{e.g.}, $32^3$) uniformly sized blocks.
Each texture block stores a particular block of the volume at a particular level in the resolution pyramid, padded by one voxel on each side to avoid bleeding from neighboring blocks during trilinear interpolation \revised{\cite{beyer2008}.}
The mapping between texture and volume blocks is maintained on the CPU.

To render a particular view of a volume, we determine a base resolution level such that screen resolution is matched for the nearest visible voxel.
Then, we prepare a 3D lookup texture in which each voxel corresponds to a volume block at base resolution.
Each voxel in this lookup texture stores the coordinates of a block in the cache texture, as well as its resolution level relative to base, encoded as a RGBA tuple.
For each (visible) volume block, we determine the optimal resolution by its distance to the viewer.
If the desired block is present in the cache texture, we encode its coordinates in the corresponding lookup texture voxel.
Otherwise, we enqueue the missing cache block for asynchronous loading through the CPU cache layer of BigDataViewer. Newly loaded blocks are inserted into the cache texture, where the cache blocks to replace are determined by a least-recently-used strategy that is also maintained on the CPU. For rendering, currently missing blocks are substituted by lower-resolution data if it is available from the cache. \revised{Intermittently, tiles may render at lower resolution or be missing completely. We prioritize maintaining interactive framerates over rendering the most complete data. Our technique is a combination of hierarchical blocking \cite{beyer2008,LaMar1999} and the missing data scheme of \cite{pietzsch2015bigdataviewer}.}

Once the lookup texture is prepared, volume rendering proceeds by raycasting and sampling volume values with varying step size along the ray, adapted to the  viewer distance.
To obtain each volume sample, we first downscale its coordinate to fall within the correct voxel in the lookup texture.
A nearest-neighbor sample from the lookup texture yields a block offset and scale in the cache texture.
The final value is then sampled from the cache texture with the accordingly translated and scaled coordinate.
With this approach, it is straightforward to raycast through multiple volumes simultaneously, simply by using multiple lookup textures.
It is also easy to mix in smaller volumes which are simply stored as 3D textures and do not require indirection via lookup textures.
To adapt to varying number and type of visible volumes, we generate shader sources dynamically at runtime.
Blending of volume and mesh data is achieved by reading scene depth from the depth buffer for early ray termination, thereby hiding volume values that are behind rendered geometry.

\subsection{Code-shader communication and reflection --- $G_3$}
\label{sec:CodeShaderCommunication}

In traditional OpenGL (before version 4.1), parameter data like vectors, matrices, etc. are communicated to shaders via uniforms, which are set one-by-one. In scenery, instead of single uniforms, Uniform Buffer Objects (UBOs) are used. UBOs lead to a lower API overhead and enable variable update rates. Custom properties defined for node classes that need to be communicated to the shader are annotated in the class definition with the \emph{@ShaderProperty} annotation, scenery picks up annotated properties automatically, and serializes them. See Listing \ref{lst:ShaderProperties} for an example of how properties can be communicated to the shader, and Listing \ref{lst:ShaderPropertiesGLSL} for the corresponding GLSL code for UBO definition in the shader. Procedurally-generated shaders can use a hash map storing these properties.

For all values stored in shader properties a hash is calculated, and they are only communicated to the GPU when the hash changes. \revised{Currently}, all elementary types (ints, floats, etc.), as well as matrices and vectors thereof, are supported.

\vspace*{-0.75\baselineskip}
\begin{lstlisting}[language=Kotlin, caption=Shader property example\label{lst:ShaderProperties}]
// Define a matrix and an integer property
@ShaderProperty var myMatrix: GLMatrix
@ShaderProperty var myIntProperty: Int
// For a dynamically generated shader: Store properties as hash map
@ShaderProperty val shaderProperties = HashMap<String, Any>()
\end{lstlisting}

\vspace*{-1.5\baselineskip}
\begin{lstlisting}[language=GLSL, caption=GLSL code example for shader properties\label{lst:ShaderPropertiesGLSL}]
layout(set = 5, binding = 0)
uniform ShaderProperties {
    int myIntProperty;
    mat4 myMatrix;
};
\end{lstlisting}

Determination of the correct memory layout required by the shader is done by our Java wrapper for the shader reflection library \emph{SPIRV-cross} and the GLSL reference compiler \emph{glslang}\footnote{\raggedright See \href{https://github.com/KhronosGroup/SPIRV-cross}{github.com/KhronosGroup/SPIRV-cross} and \href{https://github.com/scenerygraphics/spirvcrossj}{github.com/scenerygraphics/spirvcrossj} for our wrapper, \emph{spirvcrossj}.}. This provides a user- and developer-friendly API ($G_3$).

Furthermore, scenery supports shader factories --- classes that dynamically produce shaders to be consumed by the GPU --- and use them, e.g., when multiple volumetric datasets with arbitrary alignment need to be rendered in the same view.

\subsection{Custom rendering pipelines --- $G_3$}

In scenery, the user can use custom-written shaders and assign them on a per-node basis in the scene graph. In addition, scenery allows for the definition of fully customizeable rendering pipelines. The rendering pipelines are defined in a declarative manner in a YAML file, describing render targets, render passes, and their contents. Render passes can have properties that are adjustable during runtime, e.g., for adjusting the exposure of a HDR rendering pass. Rendering pipelines can be exchanged at runtime, and do not require a full reload of the renderer --- \textit{e.g.}, already loaded textures do not need to be reloaded.

The custom rendering pipelines enable the user/developer to quickly switch between different pipelines, thereby enabling rapid prototyping of new rendering pipelines. We hope that this flexibility stimulates the creation of custom pipelines, \textit{e.g.}, for non-photorealistic rendering, or novel applications, such as Neural Scene (De)Rendering \cite{Nalbach:2016wr,wu2017}.

\subsection{VR and preliminary AR support --- $G_1$}

\begin{figure}[tb]
 \centering
 \includegraphics[width=\columnwidth]{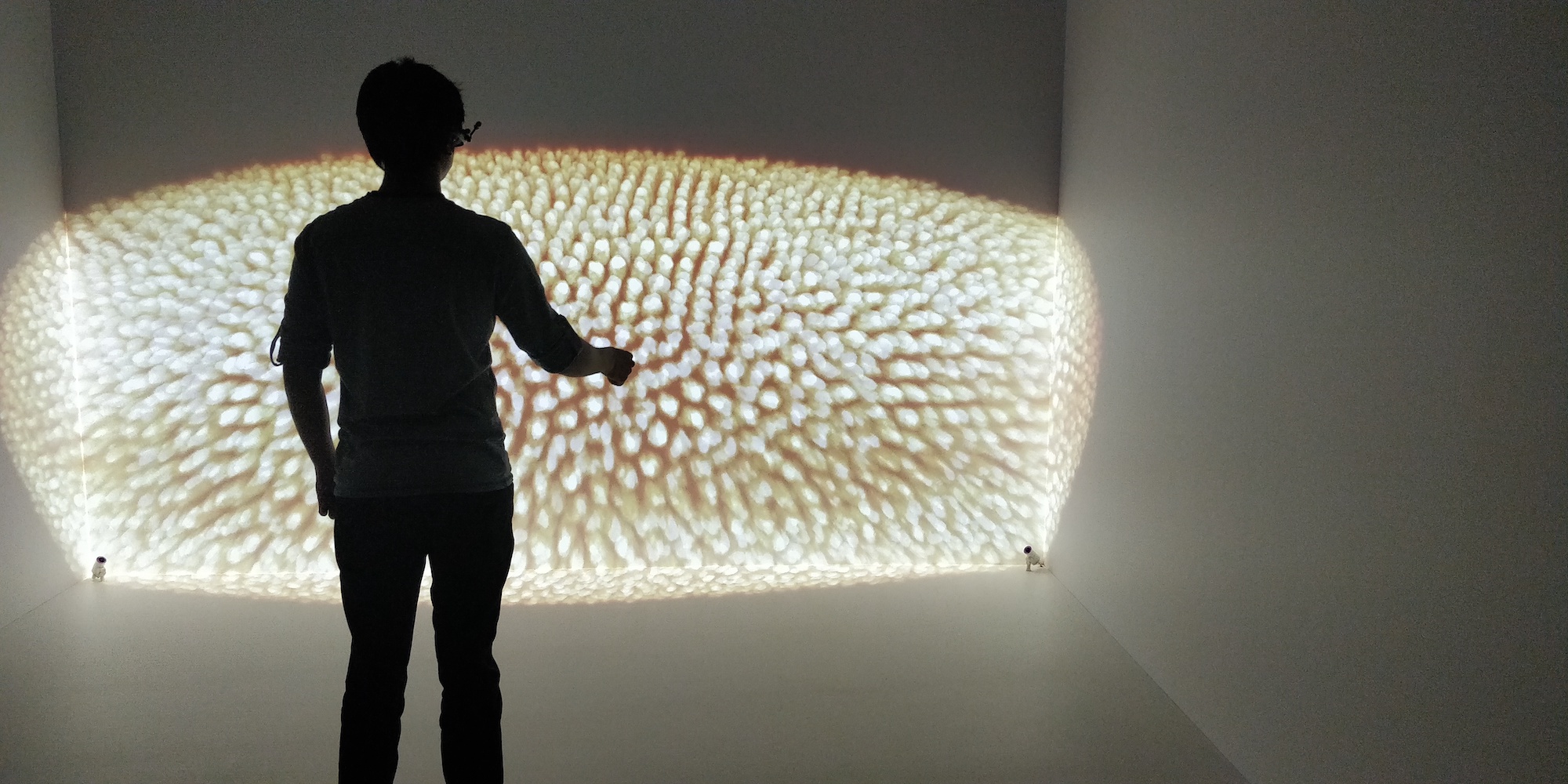}
 \caption{A scientist interactively explores a 500 GiB multi-timepoint dataset of the development of an embryo of the fruit fly \emph{Drosophila melanogaster} in the CAVE at the CSBD using a scenery-based application. Dataset courtesy of Lo\"ic Royer, MPI-CBG/CZI Biohub, and Philipp Keller, HHMI Janelia Farm \cite{Royer:2016fh}.}
 \label{fig:CAVEDrosophila}
\end{figure}

\revised{Recent reviews, e.g., \cite{Slater2016}, summarize how the use of VR/AR can lead to improved perception and navigation. Motivated by these observations,} scenery supports rendering to VR headsets via the OpenVR/SteamVR library and rendering on distributed setups, such as CAVEs or Powerwalls --- addressing $G_1$. The modules supporting different VR devices can be exchanged quickly and at runtime, as all of these implement a common interface. \revised{The use of Vulkan in turn enables improved rendering performance compared to older APIs}.

In the case of distributed rendering, one machine is designated as master, to which multiple clients can connect. We use the same hashing mechanism as described in Section~\ref{sec:CodeShaderCommunication} to determine which node changes need to be communicated over the network, use Kryo\footnote{See \href{https://github.com/EsotericSoftware/Kryo}{github.com/EsotericSoftware/Kryo}.} for fast serialization of the changes, and finally ZeroMQ for low-latency and resilient network communication. A CAVE usage example is shown in Fig.~\ref{fig:CAVEDrosophila}.

We have also developed an experimental compositor that enables scenery to render to the Microsoft Hololens.

\subsection{Remote rendering and headless rendering}

To support downstream image analysis and usage settings where rendering happens on a powerful, but non-local computer, scenery can stream rendered images out, either as raw data or as H264 stream, which can be saved to disk or streamed over the network via RTP. All produced frames are buffered and processed in a separate coroutine, such that rendering performance is not impacted.

scenery can run in headless mode, creating no windows, enabling both remote rendering on machines that do not have a screen, e.g., in a cluster setup, or easier integration testing. Most examples provided with scenery can be run automatically (see the \texttt{ExampleRunner} test) and store screenshots for comparison. In the future, broken builds will be automatically identified by comparisons against known good images.

\section{Example Applications}

\subsection{\revised{VR control of microscopes}}
\label{sec:VRControl}
\revised{We have used scenery to study VR control and visualization for state-of-the-art volumetric microscopes. In one study with 8 microscopy experts we found that users reported an improvement compared to current 2D controls, due to enhanced perception in VR.} We also investigated whether users tend to suffer from motion sickness during use of our interfaces. We found an average SSQ score \cite{kennedy1993} of $6.2\pm6.7$, which is very low, \revised{indicating that users tolerated VR rendering of live microscopy data and interaction with it well. The interface used in the study is shown in Supplementary Video 1.}

\subsection{Agent-based simulations}

\begin{figure}[tb]
 \centering
\includegraphics[width=\columnwidth]{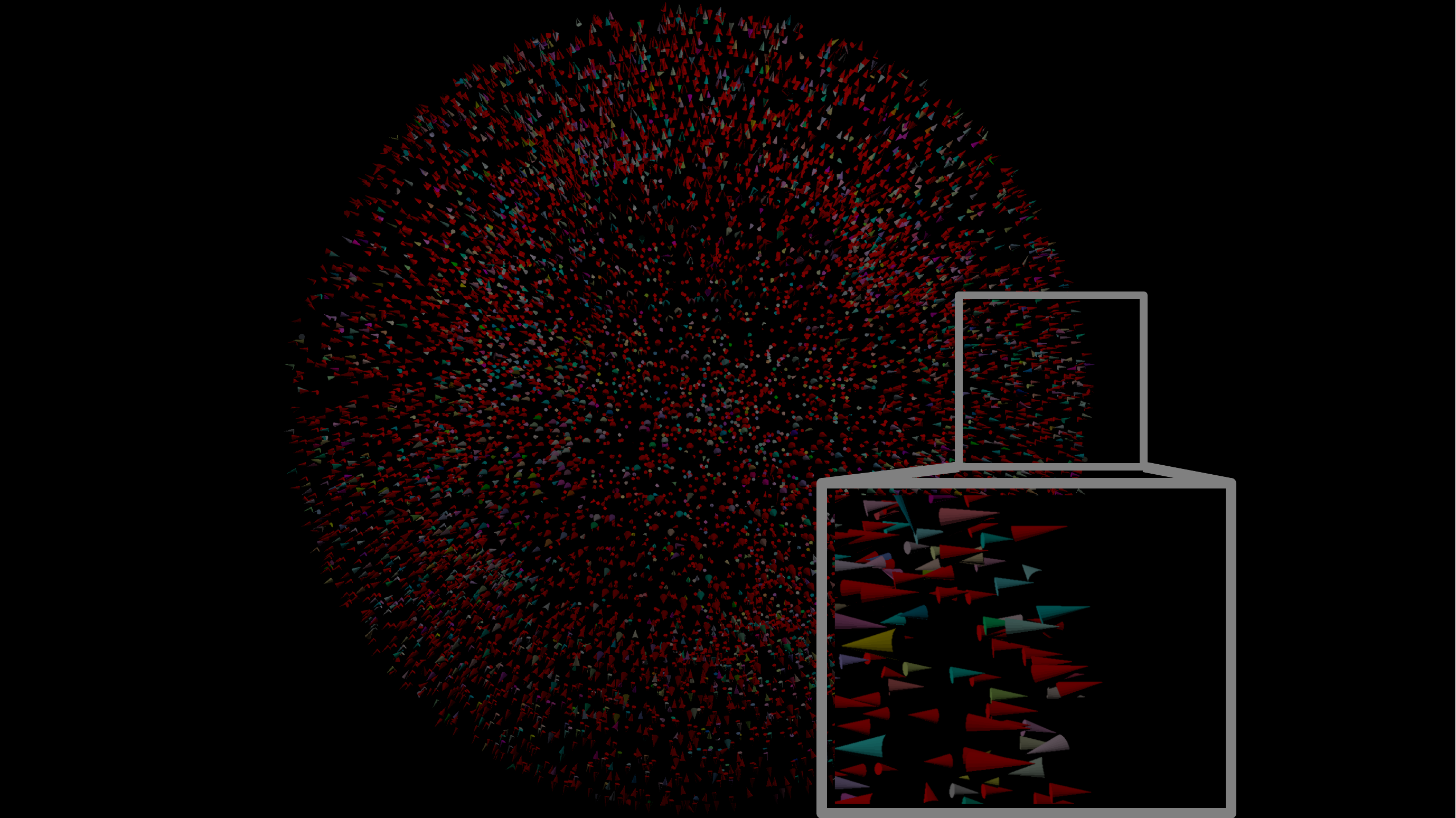}
 \caption{Agent-based simulation with 10,000 agents collectively forming a sphere.\label{fig:agentbased}}
\end{figure}

We have utilized scenery to visualize agent-based simulations with large numbers of agents. By adapting an existing agent- and physics-based simulation toolkit \cite{brevis}, we have increased the number of agents that can be efficiently visualized by a factor of 10. This performance improvement enables previous studies of swarms with evolving behaviors to be revisited under conditions that may enable new levels of emergent behavior \cite{harrington2017competitive,gold2014feedback}. In Figure~\ref{fig:agentbased}, we show 10,000 agents using flocking rules inspired by \cite{reynolds1987flocks} to collectively form a sphere.

\subsection{sciview}

\begin{figure}[tb]
 \centering
 \includegraphics[width=\columnwidth]{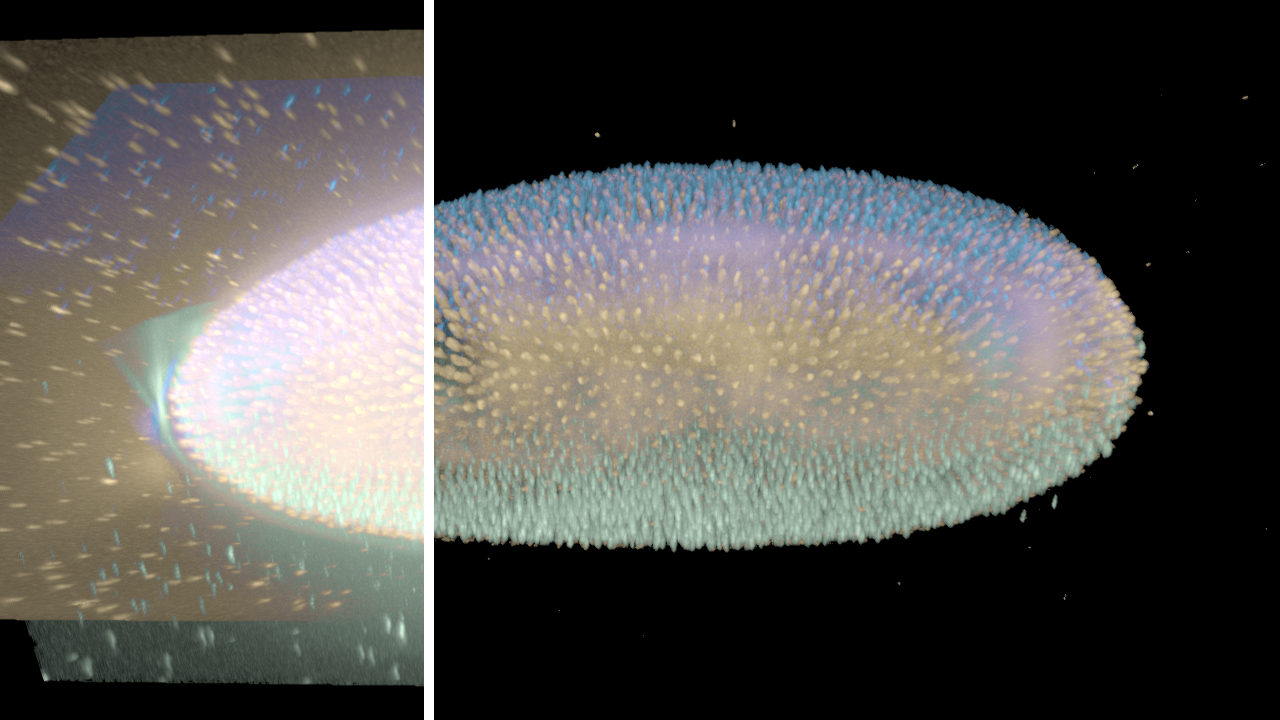}
 \caption{Out-of-core dataset of a \emph{D. melanogaster} embryo visualised with scenery/sciview. The image is a composite of three different volumetric views, shown in different colors. The transfer function on the left was adjusted to highlight volume boundaries. Dataset courtesy of Michael Weber, Huisken Lab, MPI-CBG/Morgridge Institute.\label{fig:oocdrosophila}}
 
\end{figure}

On top of scenery, we have developed a plugin for embedding in Fiji/ImageJ2 \cite{Rueden:2017ij2} --- sciview, fulfilling $G_5$. We hope it will boost the use of VR technology in the life sciences, by enabling the user to quickly prototype visualizations and add new functionality. In sciview, many aspects of the UI are automatically generated, including the node property inspector and the list of Fiji plugins and commands applicable to the currently active dataset. sciview has been used in a recent lightsheet microscopy pipeline \cite{daetwyler2019multi}. In Supplementary Video 2, we show sciview rendering  three overlaid volumes from a fruit fly embryo, a still frame of that is shown in Figure~\ref{fig:oocdrosophila}.

\section{Conclusions and Future Work}

We have introduced scenery, an extensible, user/developer-friendly rendering framework for geometric and \revised{large volumetric data} and demonstrated its applicability in several use cases. \revised{Compared to previous solutions, scenery combines the aspects of virtual reality rendering and control, with out-of-core rendering of multiple volumetric datasets in the same view, and enables the user to design their own prototypes and applications. To our knowledge, scenery is also the first framework using Vulkan on the JVM. Although scenery has undergone significant development, it is still relatively early in development compared to more mature tools and does not possess the breadth of features that are present in some alternative frameworks. However, this limitation is relaxed by compatibility with Fiji/ImageJ2, which provides a wide range of image processing capabilities.}

In the future, we will introduce better volume rendering algorithms (e.g. \cite{Kroes:2012bo, igouchkine2017}) and investigate their applicability to VR settings. Furthermore, we are looking into providing support for out-of-core mesh data, e.g. using sparse voxel octrees \cite{Kampe:2013dp,Laine:EffectiveSVO}. On the application side, we are driving forward projects in microscope control (see Section~\ref{sec:VRControl}) and VR/AR augmentation of laboratory experiments.

\section{Software and Code Availability}

scenery, its source code, and a variety of examples are available at \href{https://github.com/scenerygraphics/scenery}{github.com/scenerygraphics/scenery} and are licensed under the LGPL 3.0 license. A preview of the Fiji plugin \emph{sciview} is available at \href{https://github.com/scenerygraphics/sciview}{github.com/scenerygraphics/sciview}.

\section*{Acknowledgements}
The authors thank C. Rueden, M. Weigert, R. Haase, V. Ulman, P. Hanslovsky, W. B\"uschel, V. Leite, and G. Barbieri for additional contributions, L. Royer, P. Keller, N. Maghelli, and M. Weber for allowing use of their datasets, \revised{ and I. Tsakpinis and K. Burjack from the LWJGL community for their support. This work was supported by the European Regional Development Fund, project number CZ.02.1.01/0.0/0.0/16\_013/0001791.}

\clearpage
\bibliographystyle{abbrv-doi-hyperref-narrow}
\bibliography{scenery}

\end{document}